\documentclass[final]{aipproc}
\layoutstyle{6x9}
\usepackage{epsfig,psfig}

\newcommand{\be}{\begin{eqnarray}}
\newcommand{\ee}{\end{eqnarray}}

\def\anue{{\bar\nu_e}}

\newcommand{\dm}{\mbox{$\Delta m^2$~}}

\newcommand{\kl}{\mbox{KamLAND~}}

\newcommand{\thsol}{\mbox{$\theta_{12}$~}}

\def\ltap{\ \raisebox{-.4ex}{\rlap{$\sim$}} \raisebox{.4ex}{$<$}\ }
\def\gtap{\ \raisebox{-.4ex}{\rlap{$\sim$}} \raisebox{.4ex}{$>$}\ }
\newcommand{\sss}{\sin^2 \theta_{12}}
\def\ltap{\ \raisebox{-.4ex}{\rlap{$\sim$}} \raisebox{.4ex}{$<$}\ }
\def\gtap{\ \raisebox{-.4ex}{\rlap{$\sim$}} \raisebox{.4ex}{$>$}\ }
\newcommand{\ms}{\Delta m^2_{21}}
\newcommand{\ma}{\Delta m^2_{31}}

\newcommand{\sch}{\sin^2 \theta_{13}}
\newcommand{\dma}{\mbox{$\Delta m^2_{\rm A}$ }}

\begin{document}


\vspace*{-0.9cm}
\title{Precision measurement of oscillation parameters with reactors}

\author{Sandhya Choubey}{
  address={INFN, Sezione di Trieste, Trieste, Italy\\
Scuola Internazionale Superiore di Studi Avanzati,
I-34014,
Trieste, Italy}
}

\begin{abstract}
We review the potential of long and intermediate baseline 
reactor neutrino experiments in measuring the mass and 
mixing parameters. The KamLAND experiment can measure 
the solar mass squared difference very precisely. However 
it is not at the ideal baseline for measuring the solar neutrino 
mixing angle. If low-LMA is confirmed by the next results from 
KamLAND, a reactor experiment with a baseline of 70 km should be
ideal to measure precisely the solar neutrino mixing angle. 
If on the contrary KamLAND re-establishes high-LMA as a viable 
solution, then a 20--30 km intermediate baseline 
reactor experiment could yield 
very rich phenomenology.
\end{abstract}

\maketitle


The first results from the \kl experiment in Japan \cite{KamLAND}
has showed that the electron antineutrinos 
undergo flavor oscillations 
on their way from their source to the detector. This result coupled  
with the assumption of CPT invariance has put to rest all speculations 
regarding the true solution of the long standing solar neutrino problem, 
where the electron neutrinos produced inside the Sun apparently 
{\it disappear} as they travel from the Sun to the Earth 
(see \cite{Goswami:2003bh} for a recent review of the 
solar neutrino experiments).
This disappearance of the solar neutrinos can 
now be 
attributed confidently to neutrino flavor mixing, with the mass squared 
difference same as that relevant for the \kl experiment.
Earlier the spectacular evidence that these solar electron 
neutrinos do not really disappear, but rather appear disguised as a 
neutrino with a 
different active flavor, came from the first measurement of the 
total $^8B$ solar neutrino flux, through the neutral current 
(NC) reaction on deuterium, 
at the Sudbury Neutrino Observatory (SNO) \cite{SNO}.
The SNO NC data when combined with the data from all the other 
solar neutrino experiments picked the so called Large Mixing 
Angle (LMA) solution to the solar neutrino problem \cite{SNO,snonc}. 
This remarkable result has very recently been reinforced by the 
salt phase data from the SNO experiment \cite{sno3}. 
Prior to the SNO salt phase 
results, \kl data when combined with the other solar neutrino results, 
allowed two sub-regions within the LMA allowed region at the 99\% C.L.--
which we choose to call low-LMA (with best-fit at 
$\ms = 7.2\times 10^{-5}$ eV$^2$ and $\sss=0.3$) and high-LMA 
(with best-fit at $\ms = 1.5\times 10^{-4}$ eV$^2$ and $\sss=0.3$)
\cite{kldata}. After the SNO salt phase results,
the combined analysis 
using all available data
now allows the high-LMA solution only at the 99.13\% C.L.
\cite{salt}. Thus 
high-LMA is now further disfavored compared to low-LMA, though still not 
ruled out comprehensively.

There exists also very strong 
evidence for $\nu_{\mu} \rightarrow \nu_{\tau}$ 
($\bar{\nu}_{\mu}\rightarrow \bar{\nu}_{\tau}$) oscillations
of the atmospheric $\nu_{\mu}$ ($\bar{\nu}_{\mu}$) 
from the observed Zenith angle dependence of 
the $\mu-$like events in the 
Super-Kamiokande (SK) experiment --
with  maximal mixing and 
$1.3 \times 10^{-3} \mbox{eV}^2 \ltap
|\dma| \ltap 3.1 \times 10^{-3} \mbox{eV}^2$
($90\%$ C.L.)
\cite{SKatmo03}.

Since the solar and the atmospheric neutrino ``anomalies'' involve 
two hierarchically different mass scales, simultaneous 
description of the two 
requires the existence of three-neutrino mixing.
The solar neutrino data constrain the parameters $\Delta m^2_{\odot}
\sim  \Delta m^2_{21}$ and $\theta_{\odot} \sim \theta_{12}$ while
the atmospheric neutrino data constrain the parameters
$\Delta m^2_{atm} \sim \Delta m_{31}^2$ and $\theta_{atm} \sim \theta_{23}$.
The two sectors get connected by the mixing angle $\theta_{13}$ which
is at present constrained by the reactor data \cite{chooz}.
After these magnificent results the stage is set for the era of 
precision measurement of the oscillation parameters. 
We will discuss here 
what more can be achieved in this respect 
in reactor experiments sensitive to the 
\dm driven distortions in the 
$\anue$ spectrum due to oscillations.

The full expression for the
$\bar{\nu}_e$ survival probability in the case
of 3 flavor neutrino mixing and neutrino mass spectrum 
with normal hierarchy (NH) 
is given by \cite{3gen}
\be
P_{NH}
&& =\,~~ 1 - 2 \, \sin^2\theta_{13} \cos^2\theta_{13}\,
\sin^2\left(\frac{ \Delta{m}^2_{31} \, L }{ 4 \, E_{\nu} }\right)
- \,~~\frac{1}{2} \cos^4\theta_{13}\,\sin ^{2}2\theta_{12} \,
\sin^2\left(\frac{ \Delta{m}^2_{21} \, L }{ 4 \, E_{\nu} } \right)
\label{P21sol}  \nonumber \\
& & +\,~~ 2\,\sin^2\theta_{13}\,\cos^2\theta_{13}\, \sin^{2}\theta_{12}\,
\left(\cos
\left( \frac
{\Delta{m}^2_{31} \, L }{ 2 \, E_{\nu}} - 
\frac {\Delta{m}^2_{21} \, L }{ 2 \,
E_{\nu}}\right)
-\cos \frac {\Delta{m}^2_{31} \, L }{ 2 \, E_{\nu}} \right)\, ,
\ee
where $E_{\nu}$ is the $\bar{\nu}_e$ energy.
If the neutrino mass spectrum is
with inverted hierarchy (IH), the $\bar{\nu}_e$
survival probability can be written in the form \cite{3gen}
\be
P_{NH}
&& =\,~~ 1 - 2 \, \sin^2\theta_{13} \cos^2\theta_{13}\,
\sin^2\left(\frac{ \Delta{m}^2_{31} \, L }{ 4 \, E_{\nu} }\right)
- \,~~\frac{1}{2} \cos^4\theta_{13}\,\sin ^{2}2\theta_{12} \,
\sin^2\left(\frac{ \Delta{m}^2_{21} \, L }{ 4 \, E_{\nu} } \right)
\label{P23sol}  \nonumber \\
& & +\,~~ 2\,\sin^2\theta_{13}\,\cos^2\theta_{13}\, \cos^{2}\theta_{12}\,
\left(\cos
\left( \frac
{\Delta{m}^2_{31} \, L }{ 2 \, E_{\nu}} - 
\frac {\Delta{m}^2_{21} \, L }{ 2 \,
E_{\nu}}\right)
-\cos \frac {\Delta{m}^2_{31} \, L }{ 2 \, E_{\nu}} \right)\, ,
\ee
Thus the $\bar{\nu}_e$
survival probability
depends on the four continuous parameters 
$\ms$, $\sss$, $\ma$, $\sch$, 
and on a single ``discrete'' parameter --- the type
of the neutrino mass spectrum --- NH or IH.
Which of these parameters could be measured in a given reactor 
experiment depends on which of the terms in Eqs. (\ref{P21sol}) and 
Eqs. (\ref{P23sol}) dominate, which in turn depends 
crucially on the baseline. For a given $\Delta m^2$,
if the baseline is such that
$\sin^2(\dm L/4E) \approx 1$, the neutrinos undergo maximum flavor 
oscillations, we have a trough in the resultant $\anue$ spectrum 
and we call this a case of SPMIN (Survival Probability 
MINimum). If on the other hand the baseline corresponds to 
$\sin^2(\dm L/4E) \approx 0$, we have a peak in the $\anue$ 
spectrum and denote this case as SPMAX (Survival Probability 
MAXimum). Since the shape of the spectrum depends very crucially 
on the mass squared difference, the value of the relevant \dm can be 
expected to be determined very accurately as long as there is an 
observable distortion, irrespective of whether the distortion 
corresponds to a SPMIN or SPMAX. However since for SPMAX, the 
survival probability can be written approximately as $P_{ee}  \approx 1$, 
there is little sensitivity to $\theta$ if the SPMAX appears in the 
statistically most relevant part 
of the observed spectrum. On the other hand 
since for SPMIN the probability is 
$P_{ee} \approx 1 - \sin^2 2\theta$, one can expect maximum sensitivity 
to the mixing angle if the SPMIN is produced in the spectrum.
 
\begin{table}
\begin{tabular}{ccccc}
\hline
Data & 99\% CL &99\% CL  & 99\% CL 
& 99\% CL \cr
set & range of & spread &range  & spread  
\cr
used & $\ms\times$ & of & 
of & in  \cr
& 10$^{-5}$eV$^2$ & \dm & $\sin^2\theta_{12}$ 
& $\sin^2\theta_{12}$ \cr
\hline
only sol & 3.2 - 17.0  
& 68\% & $0.22-0.40$ & 29\% \cr
sol+162 Ty &  5.3 - 9.8 
& 30\% 
& $0.22-0.40$ & 29\%  \cr
sol+1 kTy & 6.7 - 8.0 
& 9\% & 
$0.23-0.40$ & 27\% \cr
sol+3 kTy & 6.8 - 7.6 
& 6\% & $0.24-0.37$ & 21\% \cr

\hline
\end{tabular}
\label{klbounds}
\caption
{The range of parameter values allowed at 99\% C.L.
and the corresponding spread. 
}
\end{table}

The \kl experiment in Japan is the world's first 
very long baseline reactor 
experiment which looks for disappearance of $\anue$ from nuclear 
reactors all 
over Japan. The most powerful reactors are at a distance of about 160 km. 
Thus for $\ms$ in the LMA range, this experiment 
is expected to put very stringent bounds on the allowed values 
of the solar neutrino oscillation parameters. In Table 1 we present 
the 99\% C.L. allowed range for the parameters $\ms$ and $\sss$, along 
with the corresponding \% uncertainty (``spread''), obtained by taking 
various combination of data sets into account. We note that the 
uncertainty in $\ms$ reduces from 68\% from only the solar data to 
30\% by including first \kl data. This uncertainty would further 
go down to 9\% (6\%) after 1 kTy (3 kTy) data from KamLAND.
However there seems to be little 
improvement in the uncertainty on the value of 
$\sss$, 
with increase in the \kl statistics. 

The reason for this failure of \kl to measure $\sss$ accurately enough 
can be traced back to the fact that 
the \kl spectrum shows a peak in its survival probability 
(SPMAX) at around 3.6 MeV. 
Thus, as discussed before, this sensitivity to the spectral shape 
gives \kl the ability to accurately pin down $\ms$. However since 
the oscillatory term
$\sin^2(\dm L_i/4E)$, is close to zero, it 
smothers any $\sss$ dependence along with it. Therefore,  
we conclude that \kl probably is
not at the ideal baseline for determining the solar neutrino
mixing angle.

What is the baseline most suited for measuring $\theta_\odot$?
For $\ms$ in the low-LMA region, we expect to find a 
minimum in the survival probability (SPMIN) in the 
statistically most relevant part of the 
energy spectrum when the baseline $L\sim 70$ km.
It was shown in \cite{th12} that for a
new experimental set-up with
a powerful reactor source, 
a la Kashiwazaki nuclear reactor in Japan with a maximum
power generation of about 24.6 GW, producing a SPMIN in the detected 
spectrum in a  
\kl like detector at a distance of 70 km, $\sss$ can be 
measured to within 10\% uncertainty, after 3 kTy of data. 

On the other hand, if contrary to the trend emerging in the 
solar neutrino experiments, the next \kl spectral data 
conforms to a point in the high-LMA region, then we would 
need an intermediate baseline reactor experiment with $L\sim 20-30$ km 
to get a SPMIN in the resultant spectrum. We have
shown in \cite{th12hlma} that an experimental set-up with an 
intermediate baseline of 20--30 km, a reactor with power of 5 GW and 
with 3 kTy of statistics, we can measure both $\ms$ and 
$\sss$ down to the few percent level. The impact of systematic 
uncertainties, baseline, statistics and energy threshold of the 
detector was studied \cite{th12hlma}. It was concluded that as long 
as the baseline and the energy threshold allowed the experiment 
to observe the SPMIN, $\thsol$ could be measured very accurately 
irrespective of the other conditions.

If in addition, 
the energy resolution of the detector is good enough to 
collect data in bins of 0.1 MeV width, then the intermediate baseline 
reactor experiment can observe the $\ma$ driven subdominant oscillations 
-- given by the second and the last terms in Eqs. (\ref{P21sol}) and 
(\ref{P23sol}).
Thus, this experiment can also be used very effectively to extract 
information on $\ma$, $\sch$ and even the neutrino mass hierarchy.
We have checked that an energy resolution of 
$\sigma(E)/E=5\%/\sqrt{E}$, with $E$ in MeV, 
should be good enough for this 
purpose. For $L=20$ km, bin width 0.1 MeV, systematic uncertainty 
of 2\% and 15 GWkTy statistic, one can put an upper bound of
$\sch < 0.021 (0.012)$ at $3\sigma (90\%)$ C.L.\cite{th12hlma}. 
If on the other hand 
the true value of $\sch$ turns out to be large to produce observable 
effects in this experiment, then $\ma$ can be measured to 
the percent level. For the measurement of $\sch$ and $\ma$ we 
do not necessarily need the $\ms$ to be in the high-LMA region. 
But if the condition for $\ms$ to be in high-LMA is satisfied, 
then the difference between the solar and atmospheric neutrino 
mass scales is not too severe and 
the last terms in Eqs. (\ref{P21sol}) and (\ref{P23sol}) 
are non-zero. 
Since they are different for normal (NH) and 
inverted (IH) hierarchies for $\sss$ not maximal, we can 
gain some information on the neutrino mass spectrum.
If the next \kl spectral data does bring back high-LMA as a viable 
solution, then if statistics are very high and the real value 
of $\sch$ ($\ma$) are high (low) enough, 
one can get some information on the neutrino 
mass hierarchy.
For $\ms = 1.5 \times 10^{-4}~{\rm eV^2}$, 
and statistics of $75~(125)$ GWkTy, one could
distinguish the NH from the IH spectrum
at $99.73\%$ C.L. in the region 
of $\ma \ltap 2.5 \times 10^{-3}~{\rm eV^2}$ 
if $\sin^2\theta \gtap 0.038~(0.03)$ 
\cite{th12hlma}. 

In conclusion, reactor neutrino experiments have great potential
for precision measurement of the oscillation parameters.
The KamLAND experiment can measure 
the solar mass squared difference very precisely. However 
it is not at the ideal baseline for measuring the solar neutrino 
mixing angle. If low-LMA is confirmed by the next results from 
KamLAND, a reactor experiment with a baseline of 70 km should be
ideal to measure the solar neutrino mixing angle. 
If on the contrary KamLAND re-establishes high-LMA as a viable 
solution, a 20--30 km, intermediate baseline experiment could yield 
very rich phenomenology.

\small{I acknowledge my collaborators, A. Bandyopadhyay, 
S. Goswami, S.T. Petcov and M. Piai and thank the organisers 
of the NuFact '03 workshop in New York for hospitality. 
Thanks are also due to S. Goswami for careful reading of the draft.}
\vskip -1cm

\end{document}